\documentclass[superscriptaddress,nofootinbib,12pt]{revtex4-1}
\pdfoutput=1

\makeatletter
\renewcommand{\p@subsection}{}
\renewcommand{\p@subsubsection}{}
\makeatother

\usepackage{graphicx}
\usepackage{amsmath}
\usepackage{slashed}
\usepackage{amssymb}
\usepackage[mathscr]{euscript}
\usepackage{enumerate}
\usepackage{hyperref}

\hypersetup{
	colorlinks=true,
	linktoc=page,
	linkcolor=blue,
	citecolor=blue,
	urlcolor=blue
	}

\DeclareMathOperator{\Tr}{Tr}

\newcommand{\bz}{{\bar z}}
\renewcommand{\bm}{\bar{m}}

\begin{document}

\title{Generalized gravitational entropy\\ without replica symmetry}
\author{Joan Camps}
\email{j.camps@damtp.cam.ac.uk}
\affiliation{DAMTP, Cambridge University, Wilberforce Road, Cambridge CB3 0WA, UK}
\author{William R. Kelly}
\email{wkelly@physics.ucsb.edu}
\affiliation{University of California at Santa Barbara, Santa Barbara, CA 93106, USA}
%\date{\today}

\begin{abstract}
We explore several extensions of the generalized entropy construction of Lewkowycz and Maldacena, including a formulation that does not rely on preserving replica symmetry in the bulk.  We show that an appropriately general ansatz for the analytically continued replica metric gives us the flexibility needed to solve the gravitational field equations beyond general relativity.  As an application of this observation we study Einstein--Gauss--Bonnet gravity with a small Gauss--Bonnet coupling and derive the condition that the holographic entanglement entropy must be evaluated on a surface which extremizes the Jacobson--Myers entropy.  We find that in both general relativity and Einstein--Gauss--Bonnet gravity replica symmetry breaking terms are permitted by the field equations, suggesting that they do not generically vanish.
\end{abstract}

\maketitle

\tableofcontents

\section{Introduction} \label{sec:intro}

A major goal of quantum gravity is to understand the microscopic origin of Bekenstein's formula~\cite{Bekenstein:1973ur,Bardeen:1973gs,Hawking:1974sw}
\begin{align} \label{eq:SBek}
S = \frac{\text{Area}}{4 G}.
\end{align}
One approach to studying this problem is to derive~\eqref{eq:SBek} from a path integral formulation of quantum gravity.  In the seminal paper~\cite{Gibbons:1976ue}, Gibbons and Hawking derived~\eqref{eq:SBek} for states described by a Euclidean path integral that is dominated by a $U(1)$ symmetric saddle point.

\begin{figure}
\includegraphics[width=0.4 \textwidth]{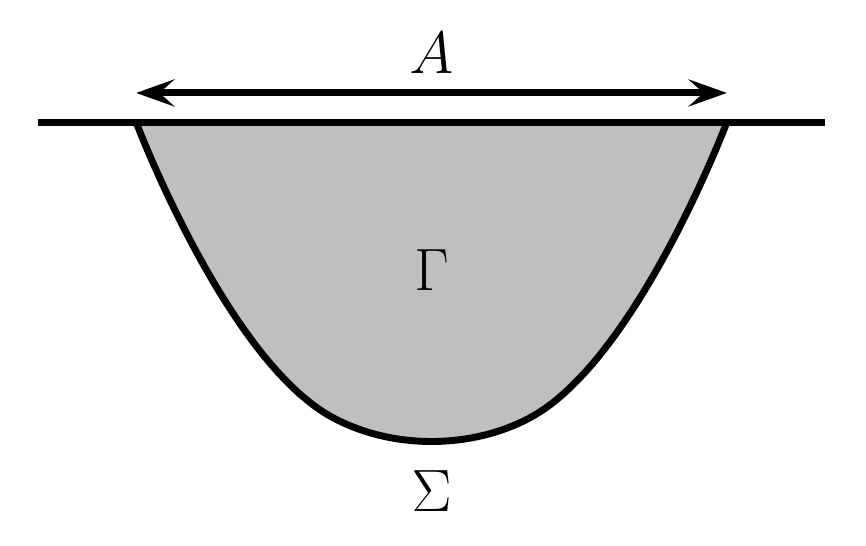} 
\caption{A sketch of the Ryu--Takayanagi surface $\Sigma$ associated with some boundary region $A$.  $\Gamma$ is a codimension-one surface satisfying $\partial \Gamma =   \Sigma \cup A$.}
\label{fig:RT}
\end{figure}

AdS/CFT has provided a comprehensive framework for understanding gravitational path integrals by identifying certain string theories with particular conformal field theories~\cite{Maldacena:1997re,Witten:1998qj,Gubser:1998bc}.  By using this correspondence~\eqref{eq:SBek} can be derived from the path integral of the dual field theory.  Ryu and Takayanagi~\cite{Ryu:2006bv,Ryu:2006ef} have proposed that this result is a special case of a more general correspondence between area and entropy.  They conjecture that in holographic theories the von Neumann entropy of the density matrix $\rho$ associated with a CFT region $A$ is given by the area of a surface in the bulk geometry, i.e.
\begin{align} \label{eq:RTformula}
S(\rho) = \frac{\text{Area}[\Sigma] }{4 G}.
\end{align}
In Euclidean AdS/CFT, the surface $\Sigma$ is defined as the minimum area codimension-two surface for which there exists a codimension-one surface $\Gamma$ satisfying $\partial \Gamma = \Sigma \cup A$ (see Fig.~\ref{fig:RT}).  This latter restriction is commonly known as the homology constraint~\cite{Headrick:2007km}.

Significant progress has been made towards deriving~\eqref{eq:RTformula} by Lewkowycz and Maldacena~\cite{Lewkowycz:2013nqa}.\footnote{see also~\cite{Fursaev:2006ih,Casini:2011kv}.}  Their derivation, which we review in section~\ref{sec:review} below, applies whenever $\Tr\rho^n$ is equal to a Euclidean path integral dominated by saddles which preserve replica symmetry.  Replica symmetry refers to a discrete global $\mathbb{Z}_n$ symmetry when the field theory path integral is computed over $n$ copies of the original manifold.  This replica construction can be used to compute the integer R\'{e}nyi entropies
\begin{align} \label{eq:Renyi}
S_n(\rho) = - \frac{1}{n-1} \log\left(\frac{\Tr[\rho^n]}{\Tr[\rho]^n}\right).
\end{align}
In AdS/CFT, $\rho^n$ is dual to a gravitational solution on a bulk manifold $M^n$ with metric $g^{(n)}$.  By analytically continuing $g^{(n)}$ to real $n$ and taking the limit $n\to1$ Lewkowycz and Maldacena calculated the von Neumann entropy and found that it is equal to the area of an extremal area surface, consistent with the formula~\eqref{eq:RTformula}.

This derivation was subsequently  extended to higher curvature theories of gravity.  Not surprisingly, several technical subtleties arise when higher curvature terms are included in the action.  Still, the Lewkowycz--Maldacena method gives a prescription for calculating the entropy functional \cite{Fursaev:2013fta, Dong:2013qoa, Camps:2013zua, Miao:2014nxa}. However, several researchers~\cite{Bhattacharyya:2013jma,Chen:2013qma,Bhattacharyya:2013gra,Bhattacharyya:2014yga} have noticed obstructions to deriving the equations of motion for $\Sigma$ when using the Lewkowycz--Maldacena ansatz for $g^{(n)}$.

This problem can be understood as follows.  In general relativity, Lewkowycz and Maldacena derive the  extremal area condition by requiring that $g^{(n)}$ satisfy the Einstein equation to leading order in $(n-1)$.  
Assuming that the matter stress tensor remans finite,
this entails discarding potentially divergent contributions to the Ricci tensor.  To first order in $(n-1)$, only the transverse-transverse components of the Ricci diverge, and these divergences can be cured by requiring that the trace of the extrinsic curvature vanish in both transverse directions.  Thus there is a precise matching between the structure of potential divergences in the field equations and the constraints necessary to fix the location of the surface $\Sigma$.  However, in higher curvature theories all components of the field equations generally diverge, and these divergences outnumber the degrees of freedom of $\Sigma$.  Furthermore, even if one focuses only on the transverse-transverse divergences, these split into ``leading'' and ``subleading", the latter suppressed by powers of $r^{n-1}$ relative to the former, where $r$ is a radial coordinate centered on the entangling surface $\Sigma$.  It was noted in~\cite{Bhattacharyya:2013gra,Dong:2013qoa} that, for a large family of higher curvature theories of gravity, requiring the leading transverse-transverse divergences to vanish extremizes the entropy on $\Sigma$.  This observation raises the question of what to do with the subleading divergences, and with the divergences in the other components.

One purpose of this paper is to resolve this problem by generalizing the Lewkowycz and Maldacena ansatz for $g^{(n)}$.  By including terms which are pure gauge for $n=1$ but physical for $n\ne 1$ we obtain a richer structure of divergences in the curvature which propagates to all components of the Ricci tensor. We will also allow $g^{(n)}$ to break replica symmetry. We present these generalizations in section~\ref{sec:manifold} and show that we are able to rederive the results of~\cite{Lewkowycz:2013nqa}.  This means showing that, despite the additional constraints from the field equations, we do not over constrain the location of $\Sigma$. The analysis also suggests that the assumption of replica symmetry can be dropped from the derivation of~\cite{Lewkowycz:2013nqa}, as we discuss below.

In section~\ref{sec:GaussBonnet} we apply our technique to general relativity plus a small Gauss--Bonnet coupling. The action for this theory is~\cite{Lanczos:1938sf,Lovelock:1971yv}
\begin{align}
I_{GB} = -\frac{1}{16\pi G} \int d^D y \, \sqrt{g} \left(  R  + \lambda (R_{\mu\nu\rho\sigma}R^{\mu\nu\rho\sigma}-4R_{\mu\nu}R^{\mu\nu}+R^2)\right) + I_\text{matter} + \dots ,
\label{eq:ActionEGB}\end{align}
where the dots indicate boundary terms and $O(\lambda^2)$ terms. More properly, the latter are  controlled by the small dimensionless parameter $\left(\lambda\, \textrm{Riem}\right)^2$. We can regard this setup as a toy model for the $\alpha^\prime$ expansions that arise in string theory~\cite{Zwiebach:1985uq}.\footnote{Because we take the coupling to be infinitesimal, our setup is free from the issues discussed in~\cite{Erdmenger:2014tba, Reall:2014pwa, Camanho:2014apa, Reall:2014sla}.}  In $D\le 4$ the term proportional to $\lambda$ is a total derivative and does not contribute to the equations of motion, so we will work in $D>4$.  It was argued in~\cite{deBoer:2011wk,Hung:2011xb} that the analog of the Ryu--Takayanagi formula~\eqref{eq:RTformula} for Einstein--Gauss--Bonnet should be the Jacobson--Myers entropy~\cite{Jacobson:1993vj,Jacobson:1993xs}
\begin{align}
S_{JM}=\frac{1}{4G}\left(\textrm{Area}+2\lambda\int_{\Sigma} d^{D-2}\sigma\sqrt{\gamma}\,\mathcal{R}\right)+\dots\,,
\label{eq:JMentropy}\end{align}
evaluated on  a surface $\Sigma$ that extremizes \eqref{eq:JMentropy}, or equivalently, on a surface which satisfies
\begin{align} \label{eq:JMeom}
\left(\gamma^{ij}-  4 \lambda\mathcal{R}^{ij}\right)\mathcal{K}_{ijz}+ O(\lambda^2) =0\,.
\end{align}
Here $\gamma_{ij}$ is the metric induced on $\Sigma$, and $\mathcal{R}_{ijkl}$ and $\mathcal{K}_{ijz}$ are its intrinsic and extrinsic curvatures.

The first half of this conjecture, namely that the appropriate entropy functional is the Jacobson--Myers entropy, was shown in~\cite{Fursaev:2013fta, Dong:2013qoa, Camps:2013zua}.  However, the derivation of the extremality condition involves accounting for the divergences mentioned above.  We find that the extra freedom afforded by our ansatz allows us to cancel all order $(n-1)$ divergences in the Einstein--Gauss--Bonnet field equations precisely when~\eqref{eq:JMeom} holds.  As in the case of general relativity, this is the only constraint on $\Sigma$.  We also find that the equations of motion allow replica symmetry breaking terms to contribute to the extrinsic curvature at $n=1$, but in a way that preserves~\eqref{eq:JMeom}.

\section{Review of the Lewkowycz--Maldacena derivation} \label{sec:review}

In this section we review the generalized gravitational entropy of \cite{Lewkowycz:2013nqa}. The purpose of the generalized entropy is to use holography to compute the von Neumann entropy
\begin{align}\label{eq:VonNeumann}
S(\rho) = - \Tr (\hat \rho \log(\hat \rho)) \, ,
\end{align}
where $\hat \rho = \rho/\Tr[\rho]$ and $\rho$ is of the form
\begin{align} \label{eq:FamilyofStates}
\rho=\mathcal{P}\left( e^{- \int_0^{2\pi} d\tau\, H(\tau) }\right)\,.
\end{align}
Here $H(\tau)$ is the Euclidean Hamiltonian and $\mathcal{P}$ indicates path ordering.  The density matrix $\rho$ can then be seen as a Euclidean time evolution operator for a time interval of length $2\pi$.

If the Hamiltonian does not depend on time then this density matrix is thermal and takes the form $\rho_T=\sum_i e^{-2\pi E_i}|E_i\rangle \langle E_i|$, in which case the considerations to follow give the usual results of black hole thermodynamics. In the remainder we will focus on the more general class of states \eqref{eq:FamilyofStates} by allowing Euclidean time-dependent features of the spacetime in the field theory side.

\begin{figure}
\includegraphics[width=0.3 \textwidth]{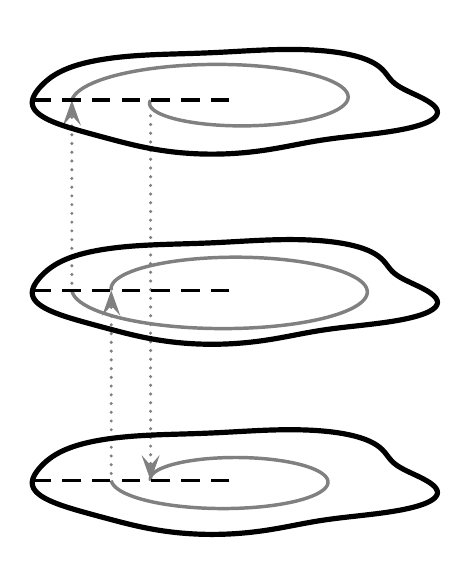} 
\caption{A sketch of the $n=3$ replica manifold.  The three solid black lines represents the $\tau$ circle of the boundary manifold $B^3$ and the dashed lines represent cuts at $\tau = 2\pi k$ for integer $k$.  The gray line is a closed curve in the bulk $M^3$ which illustrates how the three slices are glued together along the cuts.  The path integral on $B^3$ computes $\Tr[\rho^3]$ and provides a geometric realization of the formula~\eqref{eq:rhotothen}.  This path integral can also be expressed as the action associated with the metric $g^{(3)}$, a smooth metric that solves the gravitational field equation on $M^3$, as in~\eqref{eq:saddlepoint}.  Note that even if the state $\rho^3$ is replica symmetric, $g^{(3)}$ is not simply three copies of $g^{(1)}$ glued together, as the latter metric would not be smooth.
}
\label{fig:Replica}
\end{figure}

The advantage of restricting to the class of states~\eqref{eq:FamilyofStates} is that they have a geometric representation in the field theory as a path integral over some manifold $B$.  The R\'{e}nyi entropies $S_n$ (defined in~\eqref{eq:Renyi}) of these states can be written as a path integral over a manifold $B^n$ constructed by gluing together $n$ copies of the original length manifold $B$, with the trace implemented by the identification of the initial and final cuts (see Fig.~\ref{fig:Replica}):
\begin{align}\label{eq:rhotothen}
\Tr[\rho^n]=\mathcal{P}\left( e^{- \int_0^{2\pi n\sim 0} d\tau\, H(\tau) }\right)\equiv Z(n)\,.
\end{align}
The right hand side of this equation refers to the path-integral representation of this quantity as the partition function on a Euclidean manifold with $\mathbb{Z}_n$ symmetry, $B^n$. This replica symmetry is implemented by translating $\tau$ by multiples of $2\pi$: $\tau\rightarrow \tau + 2\pi s$, $s\in \mathbb{Z}/n\mathbb{Z}$. This symmetry is enhanced to $U(1)$ for the thermal state.

Holography maps these field theory calculations to a gravitational computation in one more dimension. In the semiclassical limit we have
\begin{align} \label{eq:saddlepoint}
Z(n)\approx e^{-I_n}\,,
\end{align}
where $I_n$ is the Euclidean action of a gravitational saddle point in one more dimension. We will refer to this geometry as the replica manifold $(M^n, g^{(n)})$.\footnote{Therefore we have $I_n\equiv I[g^{(n)}]$.} The field theory manifold $B^n$ is identified with the boundary of $M^n$, i.e. $\partial M^n = B^n$.  This boundary $B^n$ is $\mathbb{Z}_n$ symmetric by construction, but this symmetry need not extend into the bulk. Whether it does or not is decided dynamically.

The von Neumann entropy \eqref{eq:VonNeumann} of the state \eqref{eq:FamilyofStates} is then computed holographically as:
\begin{align}
S=-\lim_{n\rightarrow 1}\frac{1}{n-1}\log\left(\frac{e^{-I_n}}{e^{-nI_1}}\right)=\left.\partial_n\left(I_n-nI_1\right)\right|_{n=1}\,.
\label{eq:SasdifdofI}\end{align}
This expression is subtle. For one thing, it requires a prescription for analytically continuing a function defined over the positive integers $I_n$, to a function over the reals. The prescription of Lewkowycz and Maldacena \cite{Lewkowycz:2013nqa} for this continuation can be thought of as a prescription for the analytic continuation of the geometries $g^{(n)}$, whose action is $I_n$. This procedure requires, \emph{e.g}., specifying what one means by $\mathbb{Z}_n$ symmetry for non-integer $n$.  We will review this below, and see how it leads to well defined computations and familiar results for general relativity in the bulk.

The expression \eqref{eq:SasdifdofI} can be manipulated into the gravitational action of a conical singularity. To do so, start by absorbing the factor of $n$ in the second term in the right hand side into the period of Euclidean time:
\begin{align}
n I_1=n\int_0^{2\pi}d\tau\,\mathcal{L}_1=\int_0^{2\pi n}d\tau\,\mathcal{L}_1\equiv I_n[n-1]\,,
\end{align}
where the brackets indicate that we are calculating the action of a geometry with a conical excess,\footnote{We should not, however, include any contributions to $I_n[n-1]$ localized in the singularity.} of strength $2\pi(n-1)$---since we have extended the period of $\tau$. The benefit of this manipulation is that now the two geometries in the right hand side of~\eqref{eq:SasdifdofI}, the one in $I_n$ and the one in $I_n[n-1]$, have the same boundary conditions. One can therefore meaningfully compare their actions. Using the stationarity of $I_n$ 
one arrives at
\begin{align} \label{eq:Ihat}
S=\left.\partial_n \hat{I}_1[n-1]\right|_{n=1}\,.
\end{align}
The hat on $\hat{I}$ indicates the contribution to the action of an infinitesimal conical excess. To do this calculation, one first regulates the conical singularity by smoothing the tip of the cone, then calculates the action of the regulated geometry, and finally sends the regulator to zero. For general relativity, this results in $S=\textrm{Area}/4G$, in agreement with~\eqref{eq:SBek}.

If $g^{(n)}$ is replica symmetric, we can rewrite the argument in the above paragraph in terms of a conical deficit, by manipulating the first term instead of the second one in eq.\eqref{eq:SasdifdofI}, using that
\begin{align}
I_n=\int_0^{2\pi n}d\tau\,\mathcal{L}_n=n\int_0^{2\pi }d\tau\,\mathcal{L}_n=nI_1[1-n]\,.
\end{align}
However, we can not write the second equality if the replica symmetry of the boundary $\mathbb{Z}_n$ does not extend into the bulk.  Therefore, while the derivation of the holographic entanglement entropy functional as the action of a conical excess is robust against the breakdown of $\mathbb{Z}_n$, the introduction of a conical defect formally depends on $g^{(n)}$ being replica symmetric.

To derive an equation of motion for the location of the entropy surface $\Sigma$, start by noticing that if the metric $g^{(n)}$ on $M^n$ is replica symmetric, then there is a special surface $\Sigma^n$ in $M^n$ consisting of fixed points of the $\mathbb{Z}_n$ symmetry. The calculation of the entropy in terms of a conical deficit naturally localises on this surface, and the entangling surface $\Sigma$ in $M^1$ is the limit of $\Sigma^n$ as $n\to 1$.\footnote{For non-integer $n$ it no longer makes sense to identify $\tau \sim \tau + 2\pi n$. If we did, the metric $g^{(n)}$ would be discontinuous along the cut.  The prescription for computing the action in~\cite{Lewkowycz:2013nqa} is to integrate over $\tau\in [0,2\pi)$ and multiply the result by $n$. An alternative analytic continuation which does identify $\tau \sim \tau + 2\pi n$ was considered in~\cite{Prudenziati:2014tta}.} 

$\Sigma$ is however not defined by symmetry in $M^1$, as this manifold is not symmetric in general. $\Sigma$ is instead defined by an equation of motion. To find this equation of motion, consider the analytic continuation of the $\mathbb{Z}_n$--symmetric metric $g^{(n)}$ to real $n$.  A side effect of the analytic continuation is that now the Riemann tensor of $g^{(n)}$ diverges on the surface $\Sigma^{n}$.  However, requiring the equations of motion hold with a finite stress-energy tensor, i.e.
\begin{align} \label{eq:orderoflimits}
E_{\mu\nu}[g^{(n)}] = (\text{finite}) \, ,
\end{align}
where $E_{\mu\nu}$ are the gravitational field equations, results in a constraint for the location of $\Sigma^n$.  The limit $n \to 1$ of this constraint is the equation of motion for $\Sigma$.

\begin{figure}
\includegraphics[width=0.5 \textwidth]{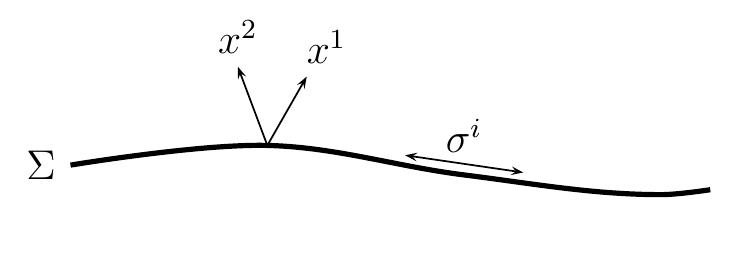} 
\caption{A sketch of the coordinates used in the text.  $\Sigma$ is the codimension-two entropy surface.  In our coordinates, $\Sigma$ is located at $x^1 = 0 =x^2$ and points on its surface are described by the $D-2$ coordinates $\sigma^i$.}
\label{fig:Coord}
\end{figure}

The key step in this argument is defining the analytically continued metric $g^{(n)}$.  To do this it is useful to introduce coordinates adapted to the surface $\Sigma^n$.  On the $D$-dimensional manifold $M^n$ let  $x^a$ for $a,b=1,2$ be transverse Cartesian coordinates to $\Sigma^n$ and let $\sigma^i$ for $i,j=3,\dots,D$ be coordinates on $\Sigma^n$ (see Fig.~\ref{fig:Coord}).  We take $\Sigma^n$ to be located at $x^1=0=x^2$.  It will also be useful to work with the polar coordinates
\begin{align}
r =\sqrt{(x^1)^2 + (x^2)^2}
, \qquad \tan\left(\frac{\tau}{n}\right) = \frac{x^2}{x^1} \, ,
\end{align}
and especially the complex coordinate
\begin{align}
z= x^1+ i x^2 = r e^{i\tau/n} , \qquad \bz = x^1- i x^2 = r e^{-i\tau/n} \, .
\end{align}
Lewkowycz and Maldacena define $g^{(n)}$ by working out  an  expansion of the metric in powers of the distance to $\Sigma^n$:
\begin{align} \label{eq:LMbc}
g^{(n)}_{\mu\nu}dy^\mu dy^\nu= dz\, d\bz+2A_{iz\bz} (\bz dz-zd\bz)d\sigma^i+\left(\gamma_{ij}+2K_{ijz}z^n+2K_{ij\bz}\bz^n\right)d\sigma^i d\sigma^j+\dots\,,
\end{align}
where the dots denote terms that  become $O(|z|^2)$ as $n\rightarrow 1$.  The metric $g^{(n)}$ is explicitly regular at integer $n$, as it contains only non-negative integer powers of the coordinates, and is invariant under $\mathbb{Z}_n$ transformations $z\rightarrow z\, e^{i\,2\pi s/n}$.  The fixed points of this replica symmetry form the codimension-two surface $\Sigma^n$, at $r=0$.

A short calculation reveals that the Riemann tensor of $g^{(n)}$ has a singularity at $\Sigma^n$ for $n\sim 1$:
\begin{align}\label{eq:RiemannGR}
R_{izjz}=-\frac{n-1}{z}K_{ijz} z^{n-1}\,,
\end{align}
which  propagates only to the $zz$ components of the Ricci tensor (and $\bz\bz$ by complex conjugation). Demanding that this singularity in the Ricci vanishes by \eqref{eq:orderoflimits}, one finds
\begin{align}
\gamma^{ij}K_{ijz}=0\,.
\label{eq:eomsminsurface}\end{align}
Upon sending $n\rightarrow1$ in the metric $g^{(n)}$, $K_{ijz}$ becomes the extrinsic curvatures of $\Sigma$, $\mathcal{K}_{ijz}$, and~\eqref{eq:eomsminsurface} becomes the equations of motion for an extremal area surface, in agreement with the Ryu--Takayangi formula. This completes our review of the generalized entropy of Lewkowycz and Maldacena.

\section{Deriving the surface equations of motion without replica symmetry} \label{sec:manifold}

In this section we will generalize the metric~\eqref{eq:LMbc} to allow for replica symmetry breaking terms as well as more general replica symmetric ones.  In section~\ref{sec:GaussBonnet} these generalizations will prove to be crucial ingredients for the solution of the field equations~\eqref{eq:orderoflimits} in Einstein--Gauss--Bonnet gravity.  Since $\Sigma^n$ is defined in~\cite{Lewkowycz:2013nqa} as the set of fixed points of the replica symmetry, part of the task of this section is to define $\Sigma$ without assuming replica symmetry.

\subsection{Defining the replica manifold} \label{subsec:defManifold}

As reviewed in section~\ref{sec:review}, the key step in Lewkowycz and Maldacena's argument is defining the analytically continued metric $g^{(n)}$.  Once the metric $g^{(n)}$ is given, the location of $\Sigma^{n}$ is restricted by the field equations~\eqref{eq:orderoflimits}.  Lewkowycz and Maldacena are able to learn what they need to know about $g^{(n)}$ and $\Sigma^n$ by assuming \eqref{eq:LMbc} and working to leading order in an expansion in powers of $(n-1)$.

We will perform the same calculation using a more general boundary condition for the surface $\Sigma^n$.  As mentioned above our boundary condition will allow $g^{(n)}$ to break replica symmetry.  For solutions which happen to be replica symmetric we can think of our calculation as a technical generalization of Lewkowycz--Maldacena, but for replica symmetry breaking $g^{(n)}$ we must supply a new definition of the surface $\Sigma^n$. 
In this case we define the metric by a boundary condition on a bulk surface which we call $\Sigma^n$.  One way to state our boundary condition is that we only allow terms which individually preserve some discrete symmetry on $M^n$ for integer $n$ (though different terms need not preserve the same discrete symmetry).  The surface $\Sigma^n$ is then the set of common fixed points of all of these discrete symmetries. This, together with regularity at integer $n$, fixes the boundary condition for $g^{(n)}$ around $\Sigma^{n}$.

We require that 
the metric near $\Sigma^n$ takes the form\footnote{In fact, it is natural to generalize~\eqref{eq:ansatz} slightly, see~\eqref{eq:newgamma} below. For the benefit of readability we postpone this discussion to section~\ref{sec:GaussBonnet}.  The solution we find for general relativity is therefore a special case of our most general ansatz in which we have set $\gamma^{(m,\bm)}_{ij} = 0$ for non-zero $m$ or $\bm$.\label{foot:gammas}}
\begin{align} \label{eq:ansatz}
 g^{(n)}_{\mu\nu} dy^\mu dy^\nu = dz\, d\bz  +\left[\hat L^{(n)}_{z\bz z} z + \hat L^{(n)}_{z\bz \bz} \bz+\text{c.c.}\right] dz\, d\bz
 + \left[(\hat L^{(n)}_{zzz} z + \hat L^{(n)}_{zz \bz} \bz) dz\, dz+\text{c.c.}\right]
 &
 \cr  
 + 2 \left[ (\hat A^{(n)}_{iz z} z  + \hat A^{(n)}_{iz \bz} \bz )dz\, d\sigma^i  +\text{c.c.}\right]
 + \left(\gamma_{ij} + \left[2 \hat K^{(n)}_{ijz}z +\text{c.c.} \right]\right) d\sigma^i d\sigma^j  \, ,&
\end{align}
where
\begin{align} \label{eq:rdependence}
\hat K^{(n)}_{ij z} =& \sum_{(m,\bm)\neq (0,0)}  z^{m(n-1)} \bz^{\bm(n-1)}     K^{(m,\bm)}_{ij z} + \dots \cr
\hat A^{(n)}_{iz z} =& \sum_{(m,\bm)\neq (0,0)} z^{m(n-1)} \bz^{\bm(n-1)}      A^{(m,\bm)}_{iz z} + \dots \cr
\hat A^{(n)}_{iz \bz} =& \sum_{(m,\bm)\geq(0,0)}  z^{m(n-1)} \bz^{\bm(n-1)}     A^{(m,\bm)}_{iz \bz} + \dots \cr
\hat L^{(n)}_{ab z} =& \sum_{(m,\bm)\neq (0,0)} z^{m(n-1)} \bz^{\bm(n-1)}     L^{(m,\bm)}_{ab z} +\dots \, .
\end{align}
Here dots denote terms that become $O(|z|)$ as $n\to 1$ and the coefficients in the expansions may depend on the $\sigma^i$.  The remaining metric functions are given by  reality conditions. Reality also implies
\begin{gather}
K^{(m,\bm)}_{ij \bz}= \bar{K}^{(\bm,m)}_{ij z}, \qquad \qquad
A^{(m,\bm)}_{i\bz\bz} = \bar A^{(\bm,m)}_{izz}, \quad 
A^{(m,\bm)}_{iz\bz} = \bar A^{(\bm,m)}_{i\bz z}, \cr
L_{\bz\bz\bz}^{(m,\bm)}=\bar{L}^{(\bm,m)}_{zzz}\,,\quad
L_{zz \bz}^{(m,\bm)} = \bar{L}_{\bz\bz z}^{(\bm,m)}\,, \quad  
L_{z\bz \bz}^{(m,\bm)}=\bar{L}_{z \bz z}^{(\bm,m)} \,.
\end{gather}
We generally use an overbar as shorthand for complex conjugation---except for $m$ and $\bm$ which are independent non-negative integers.

The boundary condition~\eqref{eq:ansatz} is explicitly regular at integer $n$ and only contains first powers of $(n-1)$.  In~\eqref{eq:rdependence} we explicitly wrote out the leading order terms in a power series about $z=0$. More precisely, we collected all terms that contribute to potential divergences in the field equations at the same rate as the singularities allowed by~\eqref{eq:LMbc}.  Note that the limits in the sums in \eqref{eq:rdependence} exclude terms that would break replica symmetry completely, as $K^{(0,0)}_{ijz}$. This would be an extrinsic curvature of $\Sigma^n$ at all integer $n$, and therefore would not preserve any subsymmetry.  Said differently, for any integer $n$ this term would be invariant under $z\rightarrow z\, e^{i\, 2 \pi/p}$ only for $p=1$. Aside from breaking replica symmetry, the main technical innovation of~\eqref{eq:ansatz} is that we have analytically continued terms which can be gauged away when $n=1$: $L_{abc}$, $A_{izz}$ and the real part of $A_{iz\bz}$.  This provides us with greater freedom to solve the equations of motion without over constraining the location of the surface $\Sigma$.

Note that of the terms appearing in~\eqref{eq:rdependence}, only the following preserve replica symmetry
\begin{align} \label{eq:RSPterms}
K^{(k+1,k)}_{ij z}\,, \qquad
A^{(k,k)}_{iz\bz}\,,\qquad
A^{(k+2,k)}_{izz}\,, \qquad
L_{z\bz z}^{(k+1,k)}\,,\qquad
L_{zzz}^{(k+3,k)}\,,\qquad
L_{zz\bz}^{(k+1,k)}\,,
\end{align}
(and their complex conjugates) for any integer $k$.  In other words, a solution that contains only these terms will be invariant under $\tau \to \tau+2\pi$ when $n$ is an integer.  All of these terms are therefore allowed when assuming replica symmetry, and we can see their inclusion as a natural generalization of the ansatz in \eqref{eq:LMbc}.

Following~\cite{Lewkowycz:2013nqa} we will solve the field equations to leading order in $(n-1)$.  However, before doing so we must specify how we will handle the factors of $z^{m(n-1)} \bz^{\bm(n-1)}$ appearing in~\eqref{eq:rdependence}.  Our prescription will be to preserve the structure of our expansion when solving the equations of motion.  For example we maintain\begin{align} 
z^{n-1} \not\sim 1+ O(n-1)\, ,
\end{align}
as well as
\begin{align} \label{eq:zfinestructure}
\frac{(n-1)}{z} (z^{n-1} - \bz^{n-1}) \not\sim O(n-1)^2  \, ,
\end{align}
even at leading order in $(n-1)$.  Keeping this structure gives us well constrained equations of motion that fix all of the terms in the power series~\eqref{eq:rdependence}.  Less restrictive conditions either give ambiguous results for the equation of motion of $\Sigma$ or allow seemingly unphysical cancellations between terms which have different angular dependence at finite $(n-1)$.\footnote{Note that~\eqref{eq:zfinestructure} would be natural if we complexified the manifold and thought of $z,\bz$ as independent coordinates, though we know of no natural reason to do so.}

Inserting the power series~\eqref{eq:rdependence} into the field equations will give us a set of constraints on the metric components.  We derive these constraints for general relativity below.

\subsection{Deriving the extremal area condition} \label{sec:Einstein}

In this section we derive the extremal area condition ${\cal K}^a=0$ for Einstein gravity using our ansatz~\eqref{eq:ansatz}.  Here ${\cal K}_{ij}{}^a$ is the extrinsic curvature of $\Sigma$ and ${\cal K}^a = \gamma^{ij} {\cal K}_{ij}{}^a$ is its trace.  Because one of our main results pertains to perturbative Einstein--Gauss--Bonnet, many of the expressions in this section will be used again in section~\ref{sec:GaussBonnet}.

Divergences only arise in the curvature of \eqref{eq:ansatz} after taking two transverse derivatives of the metric.  Thus we may write the Ricci tensor as
\begin{align} \label{eq:Ricci}
R_{ij} &= -2  \partial \bar\partial g_{ij} +\dots \cr
R_{zz} &= - \frac{1}{2} g^{ij} \partial \partial  g_{ij}  +\dots \cr
R_{iz} &= \partial\partial g_{i \bz} - \partial \bar\partial g_{i z} +\dots \cr
R_{z\bz} &=- \frac{1}{2} g^{ij} \partial \bar \partial  g_{ij} + \partial\partial g_{\bz \bz} + \bar\partial\bar\partial  g_{z z} -2 \partial\bar\partial g_{z \bz}+\dots  \, ,
\end{align}
where $\partial = \partial_z$, $\bar\partial = \partial_{\bz}$, and $\dots$ denote finite terms as $z\rightarrow 0$.  Inserting the power series expansion~\eqref{eq:rdependence} into~\eqref{eq:Ricci} gives a general expression that is conveniently expressed as
\begin{align}
R_{\mu\nu}=\sum_{m,\bm\ge0} R^{(m,\bm)}_{\mu\nu} z^{m(n-1)} \bz^{m(n-1)} \,,
\end{align}
with the following structure of divergences at the origin
\begin{subequations} \label{eq:RicciExplicit}
\begin{align}
R^{(m,\bm)}_{ij} &= - 4 (n-1) \left( \frac{\bm }{\bz} K^{(m,\bm)}_{ij z}  +\frac{m }{z} K^{(m,\bm)}_{ij \bz}   \right) \label{eq:Rij} \\
R^{(m,\bm)}_{zz} &= -  (n-1) \left(  \frac{m}{z} K^{(m,\bm)}_{ z}  -  \frac{m \bz }{z^2} K^{(m,\bm)}_{\bz}  \right) \label{eq:Rzz} \\
R^{(m,\bm)}_{iz} &= - (n-1) \left( \frac{\bm }{\bz}A^{(m,\bm)}_{izz}  - \frac{ m  }{z} (A^{(m,\bm)}_{i\bz z} -A^{(m,\bm)}_{iz\bz} ) +  \frac{m   \bz}{z^2} A^{(m,\bm)}_{i\bz \bz} \right) \label{eq:Riz} \\
R^{(m,\bm)}_{z\bz} &= \frac{\gamma^{ij}R^{(m,\bm)}_{ij}}{4} - (n-1) \left[ -\frac{m}{z} L^{(m,\bm)}_{\bz \bz z}  +\frac{2\bm }{\bz}  L^{(m,\bm)}_{z \bz z}  + \frac{\bm z}{\bz^2}L^{(m,\bm)}_{zzz} + \text{c.c.}\right]  \label{eq:Rzbz} \, ,
\end{align}
\end{subequations}
where $K^{(m,\bm)}_z = \gamma^{ij} K^{(m,\bm)}_{ij z}$, and we left implicit components that follow by complex conjugation.  The field equations demand that all of the terms in~\eqref{eq:RicciExplicit} vanish.  The constraints from~\eqref{eq:Rij} and~\eqref{eq:Rzz} are
\begin{align}  \label{eq:Keinstein}
K_z^{(m,\bm)} = 0, \qquad K^{(m,\bm\ne 0)}_{ij z} = 0 \, .
\end{align}
Note that the $K^{(m, 0)}_{ij z}$, including the leading order replica symmetric term $K^{(1, 0)}_{ij z}$, must be traceless but are otherwise unconstrained.  Next,~\eqref{eq:Riz} requires that
\begin{align}\label{eq:Aeqs}
A_{izz}^{(m,\bm \ne 0)} = 0 , \qquad A^{(m\ne 0,\bm)}_{i z \bz} = A^{(m\ne 0,\bm)}_{i \bz z}   \,.
\end{align}
Here we find that the terms $A^{(m,0)}_{izz}$ and $A^{(0,\bm)}_{i z \bz}$ are completely unrestrained.  Finally~\eqref{eq:Rzbz} and~\eqref{eq:Keinstein} imply that
\begin{align}\label{eq:Leqs}
L^{(m\neq0 ,\bm)}_{\bz\bz z} = 0 , \quad L^{(m,\bm\ne 0)}_{z\bz z} =0 , \quad L^{(m,\bm\ne 0)}_{zzz} =0  \, ,
\end{align}
which means that $L^{(m, 0)}_{z\bz z}$ and $L^{(m,0)}_{zzz}$ are unrestricted.  Note that the constraints~\eqref{eq:Keinstein},~\eqref{eq:Aeqs}, and~\eqref{eq:Leqs} do not single out replica symmetric terms in any obvious way (see~\eqref{eq:RSPterms}).

Now that we have solved the field equations in terms of $\hat K^{(n)}_{ij z}$, $\hat A^{(n)}_{iz z}$,  $\hat A^{(n)}_{i z \bz}$ and $\hat L^{(n)}_{ab z}$, we take $n\to 1$ and interpret $\hat K^{(1)}_{ij z}$, $\hat A^{(1)}_{iz z}$, $\hat A^{(1)}_{i z \bz}$ and $\hat L^{(1)}_{ab z}$ as metric functions of $g^{(1)}$.  This gives
\begin{gather} \label{eq:nfirst}
\hat K^{(1)}_{ij z} = \sum_{(m,\bm)\neq (0,0)}   K^{(m,\bm)}_{ij z}    \cr
\hat A^{(1)}_{iz z} = \sum_{(m,\bm)\neq (0,0)} A^{(m,\bm)}_{iz z}  ,\qquad
\hat A^{(1)}_{iz \bz} = \sum_{(m,\bm)\ge (0,0)} A^{(m,\bm)}_{iz \bz}  \cr
\hat L^{(1)}_{ab z} = \sum_{(m,\bm)\neq (0,0)}   L^{(m,\bm)}_{ab z}  ,\qquad
\hat L^{(1)}_{ab \bz} = \sum_{(m,\bm)\neq (0,0)} L^{(m,\bm)}_{ab \bz}  \, .
\end{gather}
Applying the constraints~\eqref{eq:Aeqs} and~\eqref{eq:Leqs} we see that $ \hat A^{(1)}_{iz z}$, $\hat A^{(1)}_{iz \bz}$ and $\hat L^{(1)}_{z\bz z}$, $\hat L^{(1)}_{zz z}$ are unrestricted by the equations of motion.  This follows immediately form the fact that $A^{(m,0)}_{izz}$, $A^{(0,0)}_{iz\bz}$, $L^{(0,\bm)}_{\bz\bz z}$, $L^{(m,0)}_{z\bz z}$,  $L^{(m,0)}_{zz z}$ are all free of constraints.  Similarly $\hat K^{(1)}_{ijz}$ is only required to satisfy $\hat K^{(1)}_z = 0$.  The form of our ansatz dictates that $\mathcal{K}_a = \hat{K}^{(1)}_a$, therefore we have
\begin{align} \label{eq:calKeq0}
{\cal K}^a=0 \,,
\end{align}
as predicted by the Ryu--Takayanagi formula~\eqref{eq:RTformula}.

\section{Generalized entropy for Einstein--Gauss--Bonnet Gravity} \label{sec:GaussBonnet}

We now compute the correction to the construction in the previous section under the addition of a perturbative Gauss--Bonnet coupling $\lambda$ in the gravitational equations of motion. 
As explained in the introduction, we choose Gauss--Bonnet corrections for technical convenience and regard \eqref{eq:ActionEGB} as a toy model for stringy $\alpha^\prime$ corrections.

We take the Lewkowycz--Maldacena replica symmetric solution~\eqref{eq:LMbc} to be the zeroth order term in a $\lambda$ expansion. To first order, there is the same possibility of breaking replica symmetry that we found in the previous section. The key ingredient for this derivation is the same as in general relativity, namely demanding absence of singularities in the gravitational field equations to linear order in $(n-1)$.

\subsection{Linearized Einstein--Gauss--Bonnet gravity}

The field equations derived from the action~\eqref{eq:ActionEGB} read 
\begin{align}
R_{\mu\nu}-\lambda H_{\mu\nu}=  (\text{finite})  \,,
\label{pertEGBeoms}
\end{align}
where the right hand side is constructed from the matter stress tensor, which is assumed to be finite, and $H_{\mu\nu}$ is defined as
\begin{align} \label{eq:Hmunu}
H_{\mu\nu}&=-2R_{\mu}{}^{\rho\sigma\xi}R_{\nu\rho\sigma\xi}+4R^{\rho\sigma}R_{\rho\mu\sigma\nu}
+4R_{\mu}{}^{\rho}R_{\nu\rho}-2R R_{\mu\nu}\cr
& \qquad  \qquad +\frac{1}{D-2} \, g_{\mu\nu}(R_{\mu\nu\rho\sigma}R^{\mu\nu\rho\sigma}-4R_{\mu\nu}R^{\mu\nu}+R^2)\,.
\end{align}
The fact that $H_{\mu\nu}$ does not contain derivatives of the Riemann tensor is the technical reason why we choose to study this and not any other correction to general relativity.

We now expand the metric in powers of $\lambda$ as
\begin{align}
g^{(n)}_{\mu\nu} = \tilde g^{(n)}_{\mu\nu} + \lambda \, \delta g^{(n)}_{\mu\nu} + O(\lambda^2),
\end{align}
where the first term $\tilde g^{(n)}_{\mu\nu}$ is the replica symmetry preserving solution of Lewkowycz and Maldacena~\eqref{eq:LMbc}.  

We must now solve
\begin{align} \label{eq:g1EOM}
R_{\mu\nu}[\delta g^{(n)}] = H_{\mu\nu}[\tilde g^{(n)}].
\end{align}
To compute $H_{\mu\nu}[\tilde g^{(n)}]$ we need to know the Riemann tensor of $\tilde{g}^{(n)}$. Expanding the metric at one order higher than in~\eqref{eq:LMbc} we obtain
\begin{align}
\tilde{g}^{(n)}_{\mu\nu}dy^\mu dy^\nu&=\left(\tilde{\gamma}_{ij}+\left[2\tilde{K}_{ijz}{} z^n+\tilde{Q}_{ijzz}z^{2n}+\tilde{Q}_{ijz\bz}z\bz+\textrm{c.c.}\right]\right)d\sigma^i d\sigma^j+2\tilde{A}_{iz\bz} (\bz dz-zd\bz)\,d\sigma^i \nonumber\\
-\frac{4}{3}&\left[\tilde{R}_{izz\bz}z^n-\textrm{c.c.}\right] (\bz dz-z d\bz) d\sigma^i+\left(dz\,d\bz-\frac{1}{3}\tilde{R}_{z\bz z\bz}(\bz dz-zd\bz)^2\right)+\dots\,.
\label{eq:LM2ndOrder}\end{align}
Here the dots stand for terms that become $O(|z|^3)$ when $n\rightarrow 1$, and $\textrm{c.c.}$ stands for complex conjugation. We introduced the object $\tilde{Q}$,\footnote{In \cite{Camps:2013zua} $Q$ is called $\dot{K}$.} with properties $\tilde{Q}_{ijab}=\tilde{Q}_{ijba}=\tilde{Q}_{jiab}$. The metric~\eqref{eq:LM2ndOrder} is explicitly replica symmetric and regular at integer $n\geq1$.

To leading order in $(n-1)$, the components of the Riemann tensor of \eqref{eq:LM2ndOrder} are\footnote{In this expansion we only keep the terms that are either finite as $n\to 1$ or proportional to $(n-1)$ but divergent as $|z|\rightarrow 0$.}
\begin{align}
R_{ij}{}^{kl} &= \tilde{\cal R}_{ij}{}^{kl} - 4 (z\bz)^{n-1} \tilde{K}_{i}{}^{[k}{}_{z} \tilde{K}^{l]}{}_{j \bz} - 4 (z\bz)^{n-1} \tilde{K}_{i}{}^{[k}{}_{\bz} \tilde{K}^{l]}{}_{j z} \cr
R_{ij k}{}^{\bz} &= z^{n-1} \tilde{R}_{ij k}{}^{\bz}  = 2 z^{n-1}\left(\tilde{\nabla}_{[i} \tilde{K}_{j] k}{}^{\bz}+\tilde{A}_{[i}{}^{\bz z} \tilde{K}_{j] k z}\right) \cr
R_{ij}{}^{z\bz} &= \tilde{F}_{ij}{}^{z\bz} - 2 (z\bz)^{n-1}\tilde{K}_{[i}{}^{kz} \tilde{K}_{j]k}{}^{\bz} \cr
R_i{}^z{}_j{}^{\bz} &= \frac{1}{2} \tilde{F}_{ij}{}^{z\bz} -\frac{1}{2}\tilde{A}_{i}{}^{z\bz}\tilde{A}_{j}{}^{z\bz}- \tilde{Q}_{ij}{}^{z\bz}+(z\bz)^{n-1} \tilde{K}_{i}{}^{k\bz} \tilde{K}_{jk}{}^{z}\cr
R_{izjz} &=-\frac{n-1}{z}\tilde{K}_{ijz} z^{n-1}+z^{2(n-1)} \tilde{K}_{i}{}^{k}{}_{z} \tilde{K}_{jkz} -z^{2(n-1)} \tilde{Q}_{ijzz}\cr
R_{izz\bz} & =z^{n-1}\tilde{R}_{izz\bz}\cr
R_{z\bz z\bz} & =\tilde{R}_{z\bz z\bz}\, .
\label{eq:GCLM}\end{align}
The remaining components are related to those above by complex conjugation and symmetries of the indices. We defined $\tilde{F}_{ijz\bz}=-\tilde{F}_{ij\bz z}\equiv\partial_i \tilde{A}_{jz\bz}-\partial_j \tilde{A}_{iz\bz}$, which is purely imaginary.  $\tilde{\cal R}_{ijkl}$ is the curvature of the metric $\tilde{\gamma}_{ij}$ on $\Sigma^n$, with covariant derivative $\tilde{\nabla}_i$. Some of the equations \eqref{eq:GCLM} become familiar Gauss-Codacci relations for $\Sigma$ upon taking the limit $n\rightarrow 1$.

We are now ready to write the source term $H_{\mu\nu}[\tilde g^{(n)}]$.  In the series expansion
\begin{align}
H_{\mu\nu}=\sum_{m,\bm\geq0} H_{\mu\nu}^{(m,\bm)} z^{m(n-1)} \bz^{\bm (n-1)}
\end{align}
the singular terms in $H_{\mu\nu}$ are given by
\begin{subequations} \label{eq:Hdivergences}
\begin{align}
H^{(2,1)}_{ij} &=4\frac{(n-1) }{\bz}\left(\tilde K_{ik\bz}\tilde R_{j}{}^{\bz k\bz}+\tilde K_{jk\bz}\tilde R_{i}{}^{\bz k\bz}-\tilde K_{ijz}\tilde R_{k}{}^{\bz k\bz}\right)-8\frac{(n-1) }{\bz}\frac{\tilde K_{kl\bz}\tilde R^{k\bz l\bz}}{D-2}\tilde \gamma_{ij} \label{eq:Hij}\\
H_{zz}^{(1,0)} &=-4\frac{(n-1) }{z}\tilde K_{ijz}\tilde{\mathcal{R}}^{ikj}{}_{k} \label{eq:Hzz10}\\
H_{zz}^{(2,1)} &=-8\frac{(n-1) }{z}\tilde K_{ijz}\left(\tilde K^{i}{}_{kz}\tilde K^{jk}{}_{\bz}+\tilde K^{j}{}_{kz}\tilde K^{ik}{}_{\bz}\right) \label{eq:Hzz21} \\
H_{z\bz}^{(2,1)} &=2 \frac{(n-1) }{\bz} \frac{D-4}{D-2} \tilde K_{ij\bz} \tilde R^{i\bz j\bz} \\
H_{iz}^{(1,1)}&= 4\frac{(n-1)}{z}(\tilde K_{ij z}\tilde R_k{}^{jkz} -\tilde K_{jk z}\tilde R_i{}^{jkz}) \, ,
\end{align}
\end{subequations}
where we defined
$\tilde{R}_{i}{}^{\bz}{}_{k}{}^{\bz}=4\lim_{n\rightarrow 1}R_{izkz}$. There are several things to note about these sources. First, we have collected only terms linear in $(n-1)$, as this is the only dependence on which we have control.  Said differently, we obtained $H_{\mu\nu}$ by squaring the Riemann tensor of the Lewkowycz--Maldacena solution~\eqref{eq:LMbc}.  However, we only calculated the Riemann to leading order in $(n-1)$, so it does not obviously make sense to include $(n-1)^2$ terms~\eqref{eq:Hdivergences}.

Note also that, as emphasized in the introduction, $H_{\mu\nu}$ generically diverges in all components, in contrast to the divergence in the Lewkowycz--Maldacena Ricci tensor coming from~\eqref{eq:RiemannGR}, which diverges only in the $zz$ (and $\bz\bz$) components. This immediately implies that to cancel all divergences in the Einstein--Gauss--Bonnet equations of motion we need more ingredients than the ones we used in section~\ref{sec:review}.

Finally, note also that  now there is more structure in the potential divergences of the equations of motion. Namely, the divergence in the $H_{zz}$ component has two sources, $H_{zz}^{(1,0)}$ and $H_{zz}^{(2,1)}$, as observed in \cite{Bhattacharyya:2013gra, Dong:2013qoa}. We will demand that these terms cancel separately as explained at the end of section~\ref{subsec:defManifold}.

\subsection{Solving the field equations}\label{sec:solvEomsEGB}

We now solve the perturbative field equations~\eqref{eq:g1EOM}.  As in the Einstein case, this involves canceling potential divergences, which results in constraints on the terms appearing in \eqref{eq:ansatz}.  As $n\rightarrow 1$, these constraints become an equation of motion for $\Sigma$---the one following from extremizing $S_{JM}$.  Obtaining an equation of motion that neither over- nor under-constrains the surface is non-trivial since, as explained in the introduction, the potential divergences outnumber the degrees of freedom of $\Sigma$.  As we shall see, the detailed structure of \eqref{eq:ansatz} is essential for this to work.

It is necessary to start by further generalizing the boundary condition~\eqref{eq:ansatz} by allowing the induced metric $\delta \gamma_{ij}$ in $\delta g^{(n)}$ to take the form
\begin{align} \label{eq:newgamma}
\delta \gamma_{ij} \to \sum_{m,\bm \ge 0} \delta \gamma^{(m,\bm)}_{ij}z^{m(n-1)}\bz^{\bm(n-1)}\, .
\end{align}
These new terms preserve replica symmetry when $m=\bm$. They are built with the first power of $(n-1)$, and are explicitly regular at integer $n$, so they are naturally allowed by the requirements of sec.~\ref{sec:manifold} (see footnote~\ref{foot:gammas} above).

Besides naturalness, there are two main uses of the generalization~\eqref{eq:newgamma}.  First,  these terms are needed to solve the equations of motion as we will see shortly.  Second, the field equations suggest that terms like those in~\eqref{eq:newgamma} might be natural beyond first order in $(n-1)$.  This is because $H_{\mu\nu}$ contains Riemann squared terms which include the square of~\eqref{eq:RiemannGR}.  Therefore $H_{\mu\nu}$ diverges like $(n-1)^2 z^{-2}$ and beyond leading order in $(n-1)$ so must $R_{\mu\nu}[\delta g]$.  The addition of the $\gamma^{(m,\bm)}_{ij}$ allows for precisely these divergences.

With the addition of these new terms, \eqref{eq:RicciExplicit} is modified as follows\footnote{The aesthetic reason for not including the $\delta\gamma^{(m,\bm)}_{ij}$ in sec.~\ref{sec:manifold} was that $\delta \gamma\cdot K$ terms generically appear in the rhs of eqs.~\eqref{eq:RicciExplicit} inside a convolution sum (and so do $\delta \gamma\cdot A$, $\delta \gamma\cdot L$ and $ \delta\gamma\cdot\gamma$). There is only one such term in \eqref{eq:RicciwithGamma} because we are perturbing \eqref{eq:LMbc}, for which the convolution collapses: of all the $\tilde{K} ^{(m,\bm)}_{ijz}$ only $\tilde{K}^{(1,0)}_{ijz}$ are non-zero, etc.}
\begin{subequations} \label{eq:RicciwithGamma}
\begin{align} 
R^{(m,\bm)}_{ij}=&-\frac{(n-1)}{z} 4m \left( \delta K^{(m,\bm)}_{ij\bz}-\delta \gamma_{k(i}^{(m,\bm-1)}\tilde K^k{}_{j)\bz}\right) 
\cr & \qquad \qquad  \qquad \qquad   \qquad  
-\frac{(n-1)}{\bz}4\,\bm\left( \delta K^{(m,\bm)}_{ijz}-\delta \gamma_{k(i}^{(m-1,\bm)}\tilde K^k{}_{j)z}\right)  \label{eq:RijGamma}\\
R^{(m,\bm)}_{zz}=&\frac{(n-1)}{2z^2}\,m\,\tilde{\gamma}^{ij}\delta \gamma^{(m,\bm)}_{ij}
-\frac{(n-1)}{z}\left(m\, \delta K^{(m,\bm)}_{z} -    \delta\gamma_{ij}^{(m-1,\bm)}\tilde{K}^{ij}{}_z\right)\cr & \qquad \qquad\qquad\qquad\qquad
+\frac{(n-1)\bz}{z^2}\,m\,\left(\delta K^{(m,\bm)}_{\bz}-\delta \gamma^{(m,\bm-1)}_{ij}\tilde K^{ij}{}_{\bz}\right)
   \label{eq:RzzGamma}\\
R^{(m,\bm)}_{iz}=&\frac{(n-1)}{z} \, m \left[ \frac{1}{2}\left(\tilde{\nabla}^j\delta\gamma_{ji}^{(m,\bm)}
+2\tilde{A}^j{}_{z\bz}\delta\gamma^{(m,\bm)}_{ij}\right)
-\left(\delta A^{(m,\bm)}_{iz\bz}-\delta A^{(m,\bm)}_{i\bz z}\right) \right]
\cr & \qquad \qquad\qquad\qquad\qquad
-\frac{(n-1)}{\bz}\bm\, \delta A^{(m,\bm)}_{izz}
-\frac{(n-1)\bz}{z^2}m\,\delta{A}^{(m,\bm)}_{i\bz\bz}
\label{eq:RizbGamma}\\
R^{(m,\bm)}_{z\bz}=&-\frac{(n-1)}{z}\left(m\,\delta K_{\bz}^{(m,\bm)}
-\frac{1}{2}\delta\gamma_{ij}^{(m,\bm-1)}\tilde{K}^{ij}{}_{\bz}\right)
\cr & \qquad \qquad\qquad\qquad\qquad
-\frac{(n-1)}{\bz}\left(\bm\,\delta K_{z}^{(m,\bm)}
-\frac{1}{2}\delta\gamma_{ij}^{(m-1,\bm)}\tilde{K}^{ij}{}_{z}\right)
+ \delta L  \label{eq:RzzbGamma}
\,
\end{align}
\end{subequations}
where we used the condition $m\,\tilde{\gamma}^{ij}\delta\gamma_{ij}^{(m,\bm)}=0$ to simplify some of the above expressions. This follows from the cancellation of the only $1/z^2$ divergence, in \eqref{eq:RzzGamma}. We also used that in the Lewkowycz--Maldacena solution $\tilde{L}_{abc}=0$, $\tilde{A}_{izz}=\tilde{A}_{i\bz \bz}=0$, $\tilde{A}_{iz\bz}=-\tilde{A}_{i\bz z}$ and $\tilde{K}_z=0$. The term $\delta L$ in \eqref{eq:RzzbGamma} means substituting the $L$ terms of \eqref{eq:Rzbz} with $L\rightarrow \delta L$.

Now we solve the field equation~\eqref{eq:g1EOM}.  Starting with the `$zz$' component we find that the cancelation of the $1/z$ divergence in the $(1,0)$ term requires
\begin{align} \label{eq:K10}
\delta K^{(1,0)}_z - \delta \gamma^{(0,0)}_{ij} \tilde K^{ij}{}_z &= 4 \tilde {\cal R}^{ij} \tilde K_{ij z} \, ,
\end{align}
and the one in the $(2,1)$ term requires
\begin{align}\label{eq:K21}
 2\delta K^{(2,1)}_z - \delta \gamma^{(1,1)}_{ij} \tilde K^{ij}{}_z =
8\tilde K_{ijz}\left(\tilde K^{i}{}_{kz}\tilde K^{jk}{}_{\bz}+\tilde K^{j}{}_{kz}\tilde K^{ik}{}_{\bz}\right) \, .
\end{align}
For all other values of $(m,\bm)$ the cancellation of this divergence gives
\begin{align} \label{eq:Kmbm}
m\, \delta K^{(m,\bm)}_z - \delta \gamma^{(m-1,\bm )}_{ij} \tilde K^{ij}{}_z &= 0 \, ,
\end{align}
while the $\bz/z^2$ divergence implies
\begin{align}\label{eommb0}
\delta K^{(m,\bm\neq0)}_z - \delta \gamma^{(m-1,\bm\neq0 )}_{ij} \tilde K^{ij}{}_z &= 0\,.
\end{align}
Note that eqs.~\eqref{eq:Kmbm} and \eqref{eommb0} are compatible and imply that, except for the $(2,1)$ component, $\delta K^{(m,\bm\neq0)}_z=0$ and $\delta \gamma^{(m-1,\bm\neq0 )}_{ij} \tilde K^{ij}{}_z=0$.

For terms with $\bm = 0$, canceling the $1/\bz$ divergence of the `$z\bz$' component demands
\begin{align}\label{eq:gammamb0}
\delta\gamma^{(m,0)}_{ij} \tilde K^{ij}{}_z = 0 \, .
\end{align}
Combining this relation with~\eqref{eq:Kmbm} gives
\begin{align}\label{eq:Kmb0}
\delta K^{(m,0)}_z = 0 \, .
\end{align}
Combining~\eqref{eq:K10} and~\eqref{eommb0}-\eqref{eq:Kmb0} gives
\begin{align} \label{eq:KgammaTrace}
\delta K_z^{(m+1,\bar m)} - \delta \gamma_{ij}^{(m,\bar m)}\tilde K^{ij}{}_z = (4 \tilde{\cal R}^{ij}\tilde K_{ij z} )  \delta^{m,0} \delta^{\bar m,0} \, .
\end{align}

Next, the $1/\bz$ divergence in the `$ij$' equation requires that
\begin{align} \label{eq:K21ij}
-\frac{4(n-1)}{\bz}\left( \delta K^{(2,1)}_{ij z} - \delta \gamma^{(1,1)}_{k(i} \tilde K^k{}_{j) z}  \right) = H^{(2,1)}_{ij} \, ,
\end{align}
which determines $\delta K^{(2,1)}_{ij z}$ in terms of $\delta \gamma^{(1,1)}_{k(i} \tilde K^k{}_{j) z}$. Note that the trace of~\eqref{eq:K21ij} would be inconsistent with~\eqref{eommb0} if not for the fact that $\tilde \gamma^{ij}H^{(2,1)}_{ij} = 0$, which can easily be seen from~\eqref{eq:Hij}.

The `$iz$' equation can be solved with a $\delta A^{(1,1)}_{iz\bz}=-\delta A^{(1,1)}_{i\bz z}$ term. The results above imply that $\delta K_z^{(m,\bm)}$ and $\delta \gamma_{ij}^{(m,\bm)}$ drop from the `$z\bz$' equation, that can be solved by $\delta L^{(2,1)}_{z\bz z}$ or $\delta L^{(2,1)}_{zz\bz}$ and their complex conjugates. These are all replica symmetric.  The explicit expressions are messy and unilluminating. 

We thus arrive at one of our main results, which is the explicit cancellation of all the divergences in the equations of motion of Einstein-Gauss-Bonnet. Again we find that replica symmetry breaking terms can be chosen to vanish, but that this choice is not mandatory.

It is now a simple matter to extract the equation of motion for the surface:
\begin{align}
\gamma^{ij} {\cal K}_{ij z} &= (\tilde \gamma^{ij} - \lambda \delta\gamma^{ij}) (\tilde K_{ij z} +\lambda \delta \hat K^{(1)}_{ij z}) \cr
&=\lambda \sum_{m,\bm \ge 0} \left(  \delta K^{(m+1,\bm)}_{z} - \delta \gamma^{(m,\bm)}_{ij} \tilde K^{ij}{}_z \right) +O(\lambda^2) \cr
&= 4 \lambda \tilde {\cal R}^{ij} \tilde K_{ijz} + O(\lambda^2) \, ,
\label{eq:pertJMeom}\end{align}
where we have used \eqref{eq:KgammaTrace} to get the third line.  Notice that many replica symmetry breaking terms were allowed to enter in $\delta g^{(n)}$, but they all canceled  
in the equation of motion. Also, the twist potential $\mathcal{A}_{iz\bz}=\hat{A}^{(1)}_{iz\bz}$ is free, as $\delta A^{(0,0)}_{iz\bz}=-\delta A^{(0,0)}_{i\bz z}$ is unconstrained. Therefore, \eqref{eq:pertJMeom} is the only physical constraint on $\Sigma$.

Comparing this result with~\eqref{eq:JMeom}, we see that we have reproduced the equation of motion conjectured by~\cite{deBoer:2011wk,Hung:2011xb}, which means that $\Sigma$ extremizes the Jacobson--Myers entropy.

\section{Discussion} \label{sec:discussion}

In this paper we explored a number of technical and conceptual generalizations of the Lewkowycz--Maldacena methodology.  One key technical insight is that terms which can be gauged away at $n=1$ can contribute divergences to the curvature at leading order in $(n-1)$.  We found that these terms are harmless in general relativity but crucial for solving the field equations in Einstein--Gauss--Bonnet gravity (and presumably all higher curvature theories).  We also explained how a ``locally replica symmetric" boundary condition could take the place of a global $\mathbb{Z}_n$ replica symmetry.  This conceptual generalization allowed us to extend our ansatz to include replica symmetry breaking terms.  This approach has lead us to a set of well defined calculations which allow us to derive the condition that $\Sigma$ extremize the entropy in general relativity and Einstein--Gauss--Bonnet gravity.  It would be suggestive if the boundary condition~\eqref{eq:ansatz} arises naturally in some dynamical context (in the spirit of the ``cosmic brane" method of~\cite{Dong:2013qoa}).  However, we leave this interesting question for future work.

Our calculations in section~\ref{sec:GaussBonnet} complete the proof started in \cite{Bhattacharyya:2013gra,Dong:2013qoa,Bhattacharyya:2014yga} that  the surfaces on which one should evaluate the entropy are those extremizing the Jacobson--Myers entropy functional, at least when the Gauss--Bonnet coupling is perturbative. We expect the method to work similarly in general Lovelock gravity. Presumably the arguments can be made non-linear in the Lovelock coupling, although such extensions of general relativity seem to always suffer from pathologies, see e.g. \cite{Camanho:2014apa}.

We have also shown that there are no obvious obstructions to relaxing the assumption of replica symmetry in Lewkowycz and Maldacena's derivation of the extremal area condition for general relativity.  We have not addressed the pressing question of whether replica symmetry is actually broken, that we intend to do elsewhere. Deciding if this is the case involves finding whether replica breaking saddles dominate the path integral.

Replica symmetry breaking saddles that could dominate the holographic calculation of entanglement entropy were discussed in \cite{Faulkner:2013yia}, which studied three dimensional general relativity in the context of $\textrm{AdS}_3/\textrm{CFT}_2$. The possibility of replica symmetry breaking was also discussed in \cite{Lewkowycz:2013nqa}.  Other interesting features of the R\'{e}nyi entropies were considered by the authors of~\cite{Belin:2013dva}, who described non-analytic behavior of $S_n$ away from $n\sim1$ by means of an instability of the hyperbolic black hole \cite{Emparan:1999gf} of \cite{Casini:2011kv}.

Replica symmetry breaking is used in condensed matter to describe spin glasses, which are frustrated systems (see \cite{Denef:2011ee} for a review). In these systems, frustration is generated by disorder originating in random impurities. It is an exciting prospect that such a dual realization of frustration may be encoded in gravity. In fact, glassy behavior has been observed in gravitational systems in \cite{Anninos:2011vn, Anninos:2011kh, Anninos:2012gk, Anninos:2013mfa} and disorder has been studied in AdS/CMT in, e.g., \cite{Adams:2011rj, Adams:2012yi, Arean:2013mta, Hartnoll:2014cua}.

\section*{Acknowledgements}

It is a pleasure to thank David Berenstein, Don Marolf and Eric Perlmutter for interesting discussions and feedback. J.C. was supported by the European Research Council grant number ERC-2011-StG 279363-HiDGR.  W.K. was supported by the National Science Foundation under grant number PHY12-05500, by FQXi grant FRP3-1338, and by funds from the University of California.

\bibliographystyle{kp}

%\bibliography{ReplicaBreaking.bib}
\bibliography{ReplicaSymmetryv5.bbl}

\end{document}